\documentstyle[prl,epsfig,aps,multicol]{revtex}
\begin{document}
\title{On the Thermodynamics of Global Optimization}
\author{Jonathan P.~K.~Doye}
\address{FOM Institute for Atomic and Molecular Physics, 
Kruislaan 407, 1098 SJ Amsterdam, The Netherlands}
\author{David J.~Wales}
\address{University Chemical Laboratory, Lensfield Road, Cambridge CB2 1EW, UK}
\date{\today}
\maketitle
\begin{abstract}
Theoretical design of global optimization algorithms can profitably utilize
recent statistical mechanical treatments of potential energy surfaces (PES's).
Here we analyze a particular method to explain its success 
in locating global minima on surfaces with a multiple-funnel structure, 
where trapping in local minima with different morphologies is expected. 
We find that a key factor in overcoming trapping is the transformation 
applied to the PES which broadens the thermodynamic transitions. 
The global minimum then has a significant probability of occupation at 
temperatures where the free energy barriers between funnels are surmountable.

\vspace{4pt}
\end{abstract}
%\pacs{02.60.Pn,61.46.+w,36.40.Ei}
%02.60.Pn  Numerical optimization
%61.46.+w  Clusters, nanoparticles, and nanocrystalline materials
%36.40.Ei  Phase transitions in clusters

\begin{multicols}{2}

Global optimization is a subject of intense current interest, since better 
optimization techniques can produce cost reductions and greater efficiency.
It is therefore important to understand the key requirements for a 
successful global optimization method, rather than proceeding on purely
empirical or intuitive grounds.

One of the difficulties associated with global optimization 
is the exponential increase in the search space with system size.
For example, the number of possible conformations of a typical protein
is so large that it would take longer
than the age of the universe to find the native state if 
the conformations were sampled randomly (Levinthal's `paradox'\cite{Levinthal}).
This problem can be more rigorously defined using computational 
complexity theory; finding the global minimum of a protein\cite{Ngo94} or 
a cluster\cite{Wille} is NP-hard.

In fact, it is easy to design surfaces that result in 
efficient relaxation to the global minimum, 
despite very large configurational spaces\cite{Zwanzig92,Zwanzig95,JD96c}.
Such surfaces are characterized by a single deep funnel\cite{BandK97} leading 
to the global minimum, and this feature may underlie
the ability of a protein to fold to the native state\cite{Leopold,Bryngel95}. 

Hence, global optimization methods should find surfaces with a single 
funnel relatively easy to tackle. The problem is 
significantly harder if there is more than one funnel,
since there are then competing relaxation pathways.
The extreme case would be when a funnel that does not lead to the global 
minimum is thermodynamically favoured down to low temperatures. 
On cooling the system would probably be trapped in the secondary funnel.
This explains why naive simulated annealing will often fail.
A number of atomic clusters bound by the Lennard-Jones (LJ) potential 
exhibit surfaces with just such a topography.
In this paper we examine in detail one such case, 
an LJ cluster with 38 atoms, to show how a recently-developed global 
optimization method is able to overcome trapping.
Our results indicate some necessary conditions the algorithm must satisfy
if it is to succeed when applied to PES's with multiple funnels.

\begin{figure}
\begin{center}
\epsfig{figure=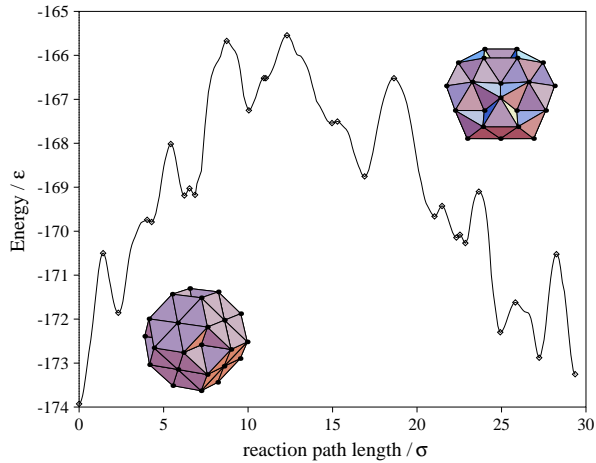,width=8.2cm}
\vglue0.1cm
\begin{minipage}{8.5cm}
\caption{Energy profile of a pathway between the two lowest energy minima 
of LJ$_{38}$, namely the fcc truncated octahedron (bottom left) 
and a structure based on the Mackay icosahedron with $C_{5v}$ 
point group symmetry (top right).
$2^{1/6}\sigma$ is the equilibrium pair separation of the LJ potential.
The method by which this pathway was obtained is described in 
Ref. \protect\cite{JD97a}.}
\end{minipage}
\label{38barrier}
\end{center}
\end{figure}

Most small LJ clusters have global minima 
based upon Mackay icosahedra\cite{Northby87}. 
However, for 38 atoms the global minimum is a face-centred cubic (fcc) 
truncated octahedron (Figure \ref{38barrier}), and for 75--77 and 102--104 
the global minima are based on Marks' decahedra\cite{Marks84}.
These minima were initially discovered by construction\cite{JD95c,JD95d}, and 
until our recent application of a `basin-hopping' algorithm \cite{WalesD97} only the LJ$_{38}$ 
global minimum had been found by an unbiased global optimization method. 

\vglue-.5cm
The multiple-funnel character of the LJ$_{38}$ PES is revealed in the energy 
profile of a pathway between the fcc global minimum and 
the lowest energy icosahedral minimum (Figure \ref{38barrier})\cite{JD97a}.
Maxima on this pathway correspond to true transition states (stationary points
with a single negative Hessian eigenvalue), and the segments linking maxima
and minima are approximate steepest descent paths.
The two lowest energy minima are well separated in configuration space,
and to cross the barrier between them
the cluster must pass through high energy 
minima associated with the liquid-like state of the cluster.
Transitions between the fcc and 
icosahedral regions of configuration space can only occur if the high energy `liquid-like' minima
have a significant probability of being occupied. At low temperatures the 
Boltzmann weights for these intermediate states are small leading to a 
large free energy barrier between the two regions.

\begin{figure}
\begin{center}
\epsfig{figure=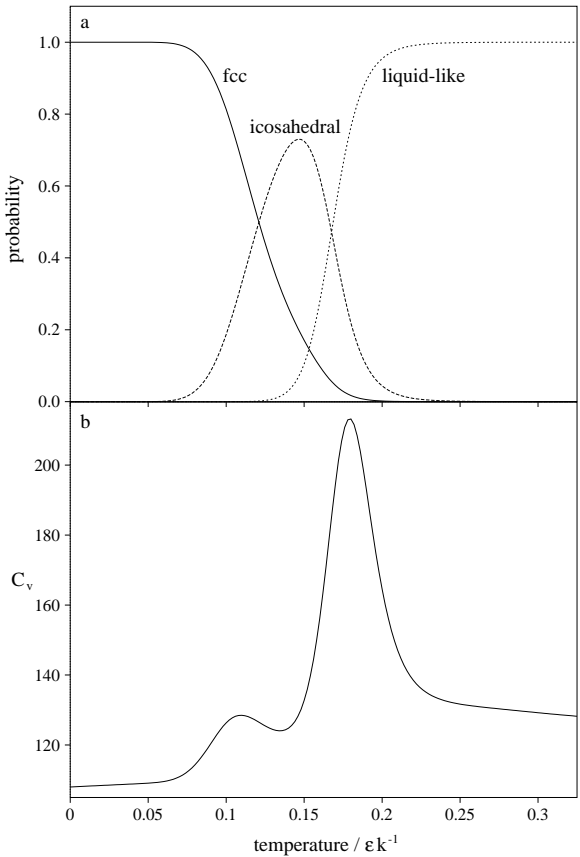,width=8.2cm}
\vglue0.1cm
\begin{minipage}{8.5cm}
\caption{Equilibrium thermodynamic properties of the 
untransformed LJ$_{38}$ PES.
(a) The probability of the cluster being in the fcc, icosahedral 
and `liquid-like' regions of bound configuration space.  
(b) The heat capacity, $C_v$.
These results were obtained by summing the anharmonic partition functions for 
a sample of minima appropriately weighted to compensate for the incompleteness 
of the sample{\protect\cite{JD95a}}.
The liquid-like region of configuration space is defined as those minima 
with $E>-171.6\,\epsilon$. }
\end{minipage}
\label{38un.therm}
\end{center}
\end{figure}
\vglue-0.3cm
Some of the equilibrium thermodynamic properties of LJ$_{38}$ are shown 
in Figure \ref{38un.therm}. 
There is only a small energy difference between the global minimum 
and the lowest energy icosahedral minimum. 
However, the entropy associated with the icosahedral structures is larger 
and icosahedra become favoured at low temperature. 
The centre of this transition is at a temperature of 0.12$\,\epsilon k^{-1}$ 
($\epsilon$ is the pair well depth of the LJ potential and 
$k$ is the Boltzmann constant); it gives rise to 
the small peak in the heat capacity (Figure \ref{38un.therm}b).
The finite-size analogue of the bulk melting transition occurs 
at 0.18$\,\epsilon k^{-1}$, producing the main peak in the heat capacity.
These transitions are also reflected in the occupation probabilities for the
different `phases' (Figure \ref{38un.therm}a).

At low temperatures one observes the cluster trapped in either 
the fcc or icosahedral funnels, 
because of the large free energy barrier between them. 
On cooling from the liquid-like state, there is a thermodynamic driving force
for entering the icosahedral region of configuration space. 
This effect is exacerbated by the topography of the PES.
The structure of simple atomic liquids has significant polytetrahedral 
character\cite{NelsonS,JD96b}, and relaxation into the icosahedral funnel
is more likely
because icosahedra have more polytetrahedral character than the fcc structures.
Hence, the icosahedral funnel is directly connected to the low energy 
liquid-like states, whereas the fcc funnel is only connected to the high 
energy liquid-like states. 
Therefore, it is extremely probable that the system will enter the icosahedral 
funnel on cooling and there become trapped. 

To observe the truncated octahedron we must simulate the 
cluster in the small temperature window where both the high energy liquid-like 
minima and the fcc structures have small, but significant, probabilities of 
being occupied.
Indeed, we did observe the truncated octahedron in this temperature range,
but only {\it once} in regular quenches from a 0.25$\,\mu$s simulation for Ar.
The situation for the larger clusters with non-icosahedral global minima 
is even worse. 
The decahedral to icosahedral transition is sharper and lies even further below 
the melting transition than for LJ$_{38}$ 
(e.g. for LJ$_{75}$ it occurs at $T=0.09\,\epsilon k^{-1}$).

The global optimization method that we analyze here belongs to the family of 
`hypersurface deformation' methods\cite{StillW88}.
In this approach the energy function is transformed, 
usually to a smoother surface with fewer minima.
The lowest minimum of this new surface is then mapped back to the original 
surface, but there is no guarantee that the global minima on the two surfaces 
are related and often there are good reasons to think that 
they are not\cite{JD95c}. 
In contrast, the transformation that we apply is guaranteed to preserve 
the global minimum. The transformed energy $\tilde E$ is defined by:
\begin{equation}
 \tilde E({\bf X}) = min\left\{ E({\bf X}) \right\}, 
\end{equation}
where ${\bf X}$ represents the vector of nuclear coordinates 
and $min$ signifies that an energy minimization is performed starting 
from ${\bf X}$.

The above approach is very similar to Li and Scheraga's Monte Carlo (MC)
minimization\cite{Li87a}. In our recent application to LJ clusters it has 
outperformed all other methods in the literature, finding all the known 
lowest energy LJ clusters up to 110 atoms, including those with
non-icosahedral global minima\cite{WalesD97}. 
The method we use to explore the $\tilde E$ surface is simply a canonical MC
simulation at constant temperature, in this case 0.8$\,\epsilon k^{-1}$.
Interestingly, the other methods that have been most successful for LJ clusters 
are genetic algorithms\cite{Deaven96,Niesse96a}
which use minimization to refine the new coordinates generated at each step,
thus in effect performing a search on the same transformed surface, $\tilde E$. 

\begin{figure}
\begin{center}
\epsfig{figure=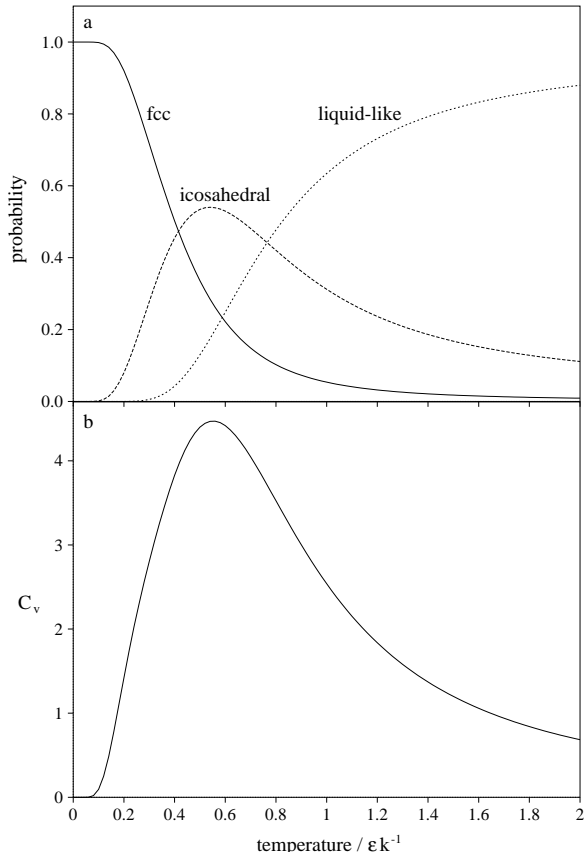,width=8.2cm}
\vglue0.1cm
\begin{minipage}{8.5cm}
\caption{Equilibrium thermodynamic properties of the transformed LJ$_{38}$ PES.
(a) The probability of the cluster being in the fcc, icosahedral 
and `liquid-like' regions of bound configuration space. 
(b) The configurational component of the heat capacity.
}
\end{minipage}
\label{38trans.therm}
\end{center}
\end{figure}
\vglue-0.3cm

The topography of the transformed surface is that of a multi-dimensional 
staircase with each step corresponding to the basin of attraction surrounding 
a minimum. 
The transformation has a significant effect on the dynamics.
Not only are transitions to a lower energy minimum barrierless, but they can 
also occur at any point along the boundary between basins of attraction, 
whereas on the untransformed surface transitions can occur only when the system 
passes along the transition state valley.
As a result intrawell vibrational motion is removed and the system can hop 
directly between minima at each step.

However, the increase in interbasin transitions will not necessarily aid the 
location of the global minimum on a multiple-funnel PES unless the 
transformation also changes the thermodynamics. 
For LJ$_{38}$ the probability of the system occupying the intermediate states 
between the funnels must be non-negligible under conditions where the 
icosahedral and fcc structures also have significant occupation probabilities. 
This is indeed what is observed for the 
$\tilde E$ surface (Figure \ref{38trans.therm}). 
The transitions have been smeared out and there is now only one 
broad peak in the heat capacity. 
Most significantly, the probability that the system is in the basin of 
attraction of the global minimum decays much more slowly, 
and the temperature range for which both the high energy liquid-like minima 
and the global minimum have significant probabilities is large.
MC simulations anywhere in this temperature range easily locate 
the global minimum from a random starting point. 
Furthermore, the $\tilde E$ transformation opens up new paths between the 
two lowest energy minima because atoms are now able to pass through oneanother. 
Such paths 
would obviously not be feasible on the untransformed PES.
Consequently, even at $T=0.2\,\epsilon k^{-1}$ the cluster can escape from the 
icosahedral region of the transformed surface.

We can understand the different thermodynamics for the two surfaces by 
writing the partition function as a sum over all the geometrically 
distinct minima on the PES. 
On the untransformed surface this gives for $p_s$, the probability 
that the cluster is in minimum $s$, 
\begin{equation}
p_s(\beta)={n_s\exp(-\beta E_s) \over
                  \overline\nu_s^{3N-6}}\big/
             \sum_s{n_s\exp(-\beta E_s) \over
                   \overline\nu_s^{3N-6}},
\end{equation}
where $\beta=1/kT$, $N$ is the number of atoms,
$\overline\nu_s$ is the geometric mean vibrational frequency of minimum $s$
(calculated by diagonalizing the Hessian matrix), 
$n_s$ is the number of permutational isomers of minimum $s$ and is given by 
$n_s=2N!/h_s$, where $h_s$ is the order of the point group of minimum $s$.
This equation is only approximate since it assumes that each well is harmonic.
However, it does give a qualitatively correct picture of the 
thermodynamics\cite{Wales93a}, and although anharmonic corrections can be 
found\cite{JD95a} they are rather cumbersome.
For the transformed surface
\begin{equation}
p_s(\beta)={n_s A_s\exp(-\beta E_s)\over
            \sum_s n_s A_s \exp(-\beta E_s)}.
\end{equation}
where $A_s$ is the hyperarea on the PES for which minimization leads to minimum $s$.
The $A_s$ values depend upon the available nuclear configuration space, which
must be bounded to prevent evaporation. 
We have considered both placing the cluster
in a container and restricting the configuration space to hyperspheres around each 
local minimum by setting the coordinates to those of the relevant minimum in the Markov chain.
Similar results are obtained for appropriate choices of container and hypersphere radii;
the $A_s$ values reported below and the results in reference \cite{WalesD97} were obtained
by resetting the coordinates.

Expressions (2) and (3) differ only in the vibrational frequency 
and $A_s$ terms. The fcc to icosahedral and the icosahedral to liquid transitions are caused by 
the greater number of minima associated with the higher temperature state, 
which leads to a larger entropy.
On the untransformed surface this effect is reinforced by the decrease in the 
mean vibrational frequencies with increasing potential energy, 
$\overline\nu_{fcc}:\overline\nu_{icos}:\overline\nu_{liquid}=1:0.968:0.864$.
Although these differences may seem small, when raised to the power $3N-6$ they 
increase the entropy of the higher energy states significantly, sharpening 
the transitions and lowering the temperature at which they occur.
In contrast, $A_s$ decreases rapidly with increasing potential energy,
$A_{fcc}:A_{icos}:A_{liquid}=1:0.0488:0.00122$.
These values were obtained by inversion of the occupation probabilities
obtained in MC runs of up to $10^6$ steps.
The decrease in $A_s$ reduces the entropy of the higher energy states, 
causing the transitions to be broadened and the temperature 
at which they occur to increase.

We can now explain in detail why 
the basin-hopping method is successful.
First, the staircase transformation removes the barriers between minima 
on the PES without changing the identity of the global minimum, 
accelerating the dynamics. 
Second, it changes the thermodynamics of the surface, broadening the transitions
so that even for a multiple-funnel surface such as that of LJ$_{38}$, 
the global minimum has a significant probability of occupation at temperatures
where the free energy barrier for passage between the funnels is surmountable.

It is this latter feature which is especially important in overcoming 
the difficulties associated with multiple funnels and represents a 
{\it new criterion for designing successful global optimization methods\/}. 

The broadened transition that results from the transformation also differs 
markedly from the optimum conditions for protein folding, 
if we assume that proteins have evolved single-funnel surfaces 
in order to fold efficiently.
A steep funnel provides a large thermodynamic driving force 
for relaxation to the global minimum\cite{JD96c,Bryngel95}, 
and also leads to a sharp thermodynamic transition. 
However, in global optimizations one cannot make assumptions about the 
topography of the PES and on a multiple-funnel surface such features can exacerbate
the difficulties associated with trapping in secondary funnels\cite{JD96c}.

Moreover, in protein folding there is the extra requirement that the folded 
protein must remain localized in the native state, and this necessitates a 
sharp transition. 
There is no need for a global optimization method to mimic this 
property. Indeed, when applied to a PES with multiple funnels
the optimum temperature for the basin-hopping approach 
is above the transition, where the system only spends a minority of 
its time in the global minimum.

We have tested the basin-hopping method on a number of other systems, 
and its performance is equally impressive. 
For example, it succeeds in finding the global minima for clusters
bound by short range Morse potentials, which have much rougher energy landscapes
than LJ clusters\cite{JD95c}. We have also obtained results for
water, metal and silicon clusters which are helping to interpret experimental results.

D.J.W.\ is grateful to the Royal Society for financial support.
The work of the FOM Institute is part of the
scientific program of FOM and is supported by the Nederlandse
Organisatie voor Wetenschappelijk Onderzoek (NWO).

% figures follow here

\end{multicols}
\end{document}